\documentclass[iop]{emulateapj}
\shorttitle{$\omega$ Cen IMBH X-ray Limit}
\shortauthors{Haggard {\it et al.}}

\usepackage{graphicx}
\usepackage[fleqn]{amsmath}

\def\wcen{$\omega$~Cen}

\def\lbol{${L}_{\rm bol}$}
\def\ledd{${L}_{\rm Edd}$}

\def\lx{${L}_{\rm X}$}
\def\fx{${f}_{\rm X}$}

\def\rc{r$_c$}
\def\mass{${\cal M}$}
\def\msun{${\cal M}_{\odot}$}
\def\mdot{$\dot{\cal M}$}
\def\ergs{erg s$^{-1}$}
\def\ergscm2{erg s$^{-1}$ cm$^{-2}$}
\def\yr-1{yr$^{-1}$}

\def\x{$\times$}

\def\simlt{\buildrel{<}\over \sim}
\def\simgt{\buildrel{>}\over \sim}

\def\simlt{$\la$}
\def\simgt{$\ga$}

\def\asec{$''$}
\def\amin{$'$}
\def\secspt{$\buildrel{\prime\prime}\over .$}
\def\minspt{$\buildrel{\prime}\over .$}

\def\HST{$\it HST$}
\def\Chandra{$\it Chandra$}

\submitted{Accepted to to ApJ Letters on July 20, 2013}

\begin{document}

\title{A Deep Chandra X-ray Limit on the Putative IMBH in Omega Centauri}

\author{Daryl Haggard,\altaffilmark{1}
Adrienne M. Cool,\altaffilmark{2}
Craig O. Heinke,\altaffilmark{3}
Roeland van der Marel\altaffilmark{4}
Haldan N. Cohn,\altaffilmark{5}
Phyllis M. Lugger,\altaffilmark{5} and
Jay Anderson\altaffilmark{4}}

\altaffiltext{1}{CIERA Fellow, Center for Interdisciplinary Exploration and Research in Astrophysics, Physics and Astronomy Department, Northwestern University, 2145 Sheridan Road, Evanston, IL 60208, USA; dhaggard@northwestern.edu}
\altaffiltext{2}{Department of Physics and Astronomy, San Francisco State University, 1600 Holloway Ave., San Francisco, CA 94132, USA; cool@sfsu.edu}
\altaffiltext{3}{Department of Physics, University of Alberta, Room 238 CEB, Edmonton, AB T6G 2G7, Canada}
\altaffiltext{4}{Space Telescope Science Institute, 3700 San Martin Drive, Baltimore, MD 21218, USA}
\altaffiltext{5}{Department of Astronomy, Indiana University, 727 E. Third St., Bloomington, IN 47405, USA}

\begin{abstract}

We report a sensitive X-ray search for the proposed intermediate mass black hole (IMBH) in the massive Galactic cluster, $\omega$ Centauri (NGC 5139). Combining {\it Chandra X-ray Observatory} data from Cycles 1 and 13, we obtain a deep ($\sim291$ ks) exposure of the central regions of the cluster. We find no evidence for an X-ray point source near any of the cluster's proposed dynamical centers, and place an upper limit on the X-ray flux from a central source of \fx($0.5-7.0$ keV) $\le 5.0 \times10^{-16}$ erg cm$^{-2}$ s$^{-1}$, after correcting for absorption. This corresponds to an unabsorbed X-ray luminosity of \lx($0.5-7.0$ keV) $\le 1.6 \times10^{30}$ erg s$^{-1}$, for a cluster distance of 5.2 kpc, Galactic column density N$_{\rm H} = 1.2\times10^{21}$ cm$^{-2}$, and powerlaw spectrum with $\Gamma = 2.3$. If a $\sim10^4$ \msun\ IMBH resides in the cluster's core, as suggested by some stellar dynamical studies, its Eddington luminosity would be \ledd\ $\sim10^{42}$ erg s$^{-1}$. The new X-ray limit would then establish an Eddington ratio of \lx/\ledd\ \simlt\ $10^{-12}$, a factor of $\sim$10 lower than even the quiescent state of our Galaxy's notoriously inefficient supermassive black hole Sgr~A*, and imply accretion efficiencies as low as $\eta$ \simlt\ $10^{-6}-10^{-8}$. This study leaves open three possibilities: either \wcen\ does not harbor an IMBH or, if an IMBH does exist, it must experience very little or very inefficient accretion.

\end{abstract}


\section{Introduction}

\begin{figure*}[th]
\vspace{0.8in}
\plottwo{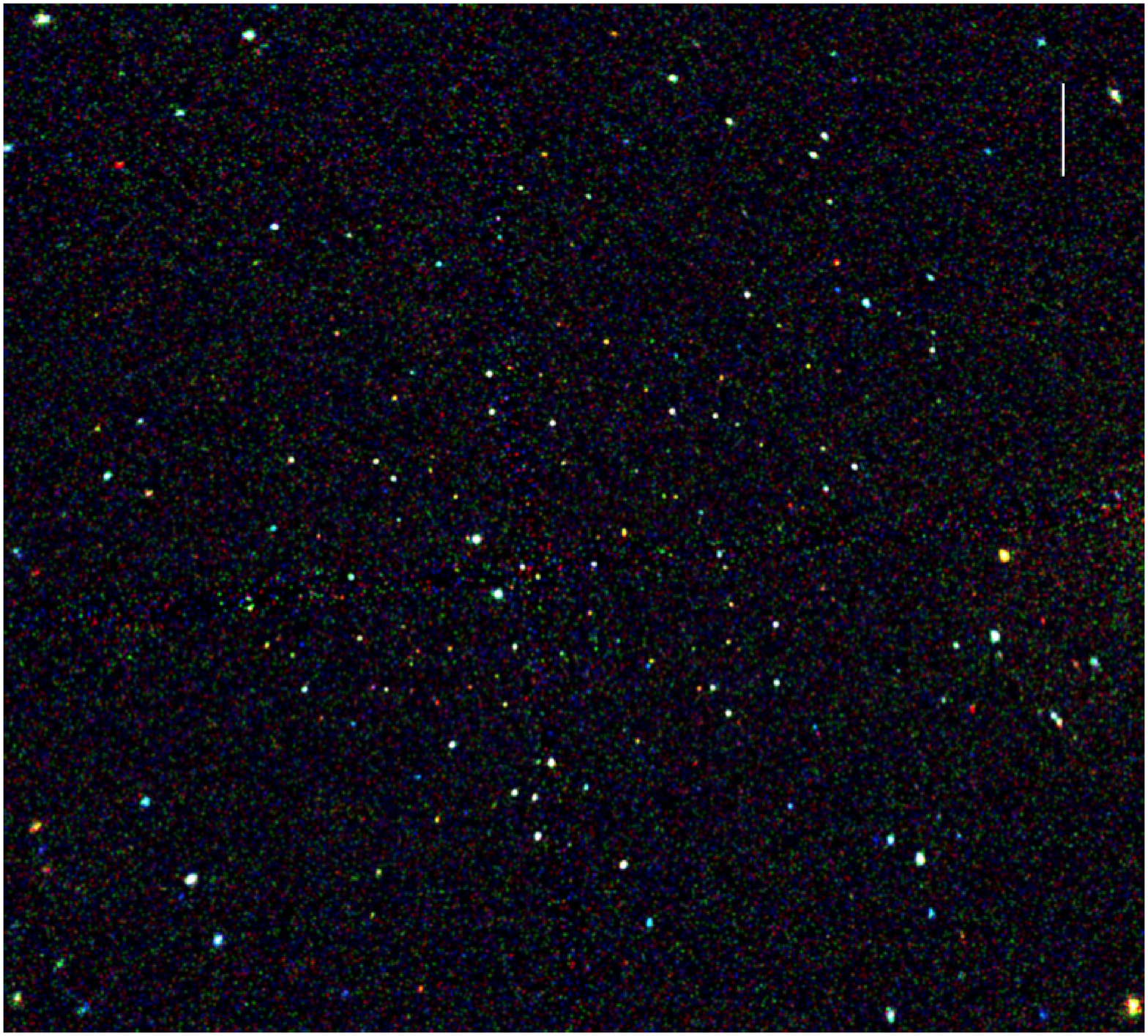}{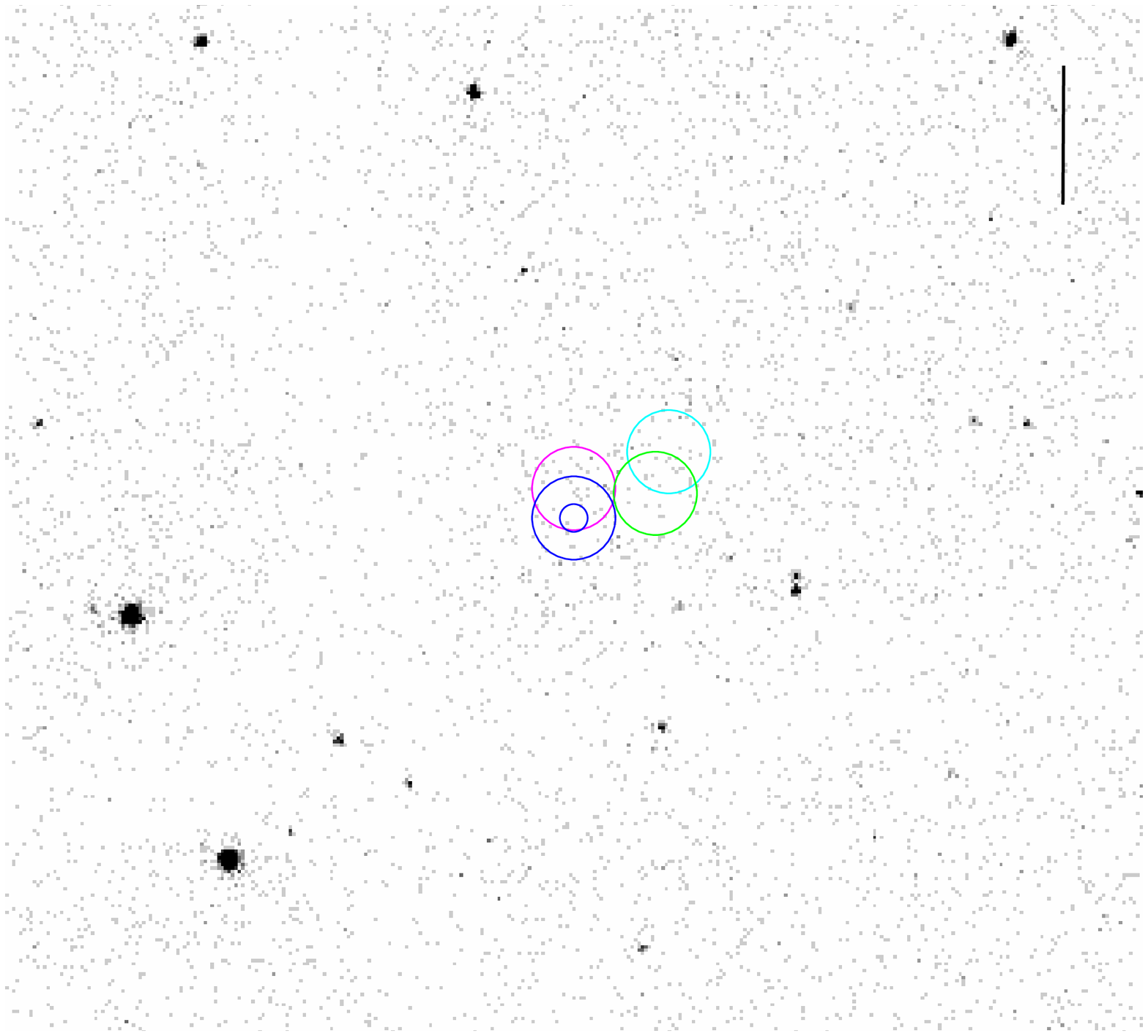}
\figcaption{({\it Left}\,) \Chandra\ three-color X-ray image of the core of \wcen\ (red: 0.5--1.2 keV, green: 1.2--2.0 keV, blue: 2.0--7.0 keV). The field of view is approximately 12\amin\ \x\ 11\amin\ ($\sim$2.5 core radii [\rc\ $\sim$ 2.6\amin]; white scale bar: 1\amin); North is up and East is to the left. The total combined \Chandra\ exposure time is 290.9 ks. ({\it Right}\,) Broad band ($0.5-7.0$ keV) image of the inner 2\minspt7 $\times$ 2\minspt5 (black scale bar: 20\asec) --- no X-ray source is detected at or near any of the cluster's proposed centers (Table \ref{tab_wcen}). The small (2\asec radius) blue circle marks the AvdM10 dynamical center; the large 6\asec\ radius circles indicate our search areas at the AvdM10 (blue), \citet{Noyola10} (magenta), \citet{Noyola08} (green), and \citet{Harris96} (cyan) centers. These represent the regions within which the putative IMBH may have ``wandered'' due to Brownian motion. The brightest X-ray sources are cataclysmic variables identified in \HST\ images \citep{Cool13}, allowing for accurate placement of the optical center on the image (see \S\ref{obs} for details). \\
(A color version of this figure is available in the online journal.)} 
\label{fig_wcen_data}
\end{figure*}

Do 
globular clusters (GCs) harbor intermediate-mass black holes (IMBHs; $\sim10^2-10^6$\msun)? Despite nearly four decades of study \citep[e.g.,][]{Bahcall75}, this fundamental question in black hole astrophysics remains unanswered. IMBHs making up $\sim$0.1--1\% of a cluster's mass could have formed at early times from runaway mergers of massive stars \citep{PortegiesZwart04}, or from the evolution of population III stars \citep{Madau01}. If they exist, these black holes would be of significant astrophysical interest, with potential connections to the assembly of supermassive black holes and to the first stars in the Universe \citep[e.g.,][]{Volonteri12}.

In the last decade, many GCs have been systematically searched for IMBHs. Dynamical measures of stars in the central regions have been undertaken to reveal the IMBH's influence, e.g., via high velocity dispersions near the sphere of influence \citep[e.g.,][and references therein]{Gebhardt00} and have resulted in the identification of several candidates. Yet, these measurements are complicated both by competing physical mechanisms, e.g., mass segregation of populations of lower-mass compact objects \citep[e.g.,][]{Illingworth76,Baumgardt03}, and by limitations in the available data, e.g., shot noise due to small numbers of stars within the sphere of influence \citep[e.g.,][vdMA10 hereafter]{vdMA10}. 

Searches for accretion signatures at X-ray and radio wavelengths provide a complementary approach. G1, arguably the most convincing GC IMBH candidate in the Local Group \citep{Ulvestad07}, has recently been studied by \citet{MillerJones12} with simultaneous X-ray and radio observations. They find X-ray emission consistent with earlier observations, but no detectable radio continuum emission, and argue that previous radio detections arise from flaring activity from a black hole low-mass X-ray binary (LMXB). As part of a larger campaign with the {\it Jansky Very Large Array} (JVLA), \citet{Strader12} studied the cores of M15, M19, and M22, but uncovered no point sources consistent with the clusters' centers. They place 3$\sigma$ upper limits on the IMBH masses of $360-980$ \msun. Thus, no clear evidence for IMBHs in globular clusters exists, even as there is substantial support for IMBHs in small galaxies \citep[e.g.,][]{Greene07a,Thornton08}.

As the largest globular cluster in the Milky Way \citep[$3\times10^6$\msun,][]{Meylan02}, or the possible remnant of an accreted dwarf galaxy \citep{Norris96}, \wcen\ is a prime candidate to harbor an IMBH. Significant efforts have been undertaken to search for one. \citet{Noyola08,Noyola10} have claimed the dynamical detection of a $\sim$40,000\msun\ IMBH based on {\it Hubble Space Telescope} (\HST) data and integral field unit spectroscopy from {\it Gemini} and {\it VLT}. However, another study based on \HST\ proper motions \citep[vdMA10;][hereafter AvdM10]{Anderson10}, achieves dynamical fits that do not require a massive compact central object, though their $1.8\times10^{4}$\msun\ (3$\sigma$) upper limit still leaves room for a black hole in the IMBH mass range.

Most recently, \citet{Lu11} have found a 3$\sigma$ peak in their 5.5 GHz radio map of \wcen\ at a position consistent with the center measured by AvdM10. Deeper radio imaging is required to determine if this peak is real, but the discovery of such peaks at the centers of both \wcen\ and 47 Tuc is notable \citep{Lu11}. 
 
{\it ROSAT} observations of \wcen\ have revealed multiple sources in the cluster core, but none coincident with the cluster center \citep{Verbunt00}. Cycle 1 \Chandra\ observations, sensitive to sources as faint as \lx($0.5-2.5$ keV) $\approx 1\times10^{30}$ \ergs\ \citep{Haggard09}, revealed many more sources in the core, but again nothing at the cluster center. 

We have recently acquired Cycle 13 \Chandra\ observations that increase the total exposure by a factor of \simgt four. We present results of a more sensitive IMBH search carried out using the full \Chandra\ data set. In Section 2, we describe the observations and our search for sources in and near the cluster center. 
In \S 3, we present new IMBH X-ray flux, luminosity, and mass limits, and discuss the implications of this non-detection; in \S 4 we summarize our findings.

\section{Observations and Analysis}
\label{obs}

We obtained four \Chandra\ exposures of \wcen\ using the imaging array of the Advanced CCD Imaging Spectrometer (ACIS-I) in ``very faint'' mode, on 2000 January 24--25 (ObsIDs 653, 1519) and 2012 April 16--17 (ObsIDs 13726, 13727). We reduced and analyzed the data using CIAO\footnote{\Chandra\ Interactive Analysis of Observations (CIAO) version 4.5 and Calibration DataBase (CALDB) version 4.5.5.1 \citep{Fruscione06}.}, reprocessing the observations and combining them using the {\tt chandra\_repro} and {\tt merge\_obs} scripts. The total combined on-axis exposure time is 290.9 ks. We used the known \HST\ positions for the two brightest core sources \citep[CVs 13a and 13c,][]{Haggard09,Cool13} to perform a boresight correction ($\Delta {\rm RA}, \Delta {\rm Dec} = $ 0\secspt0037, $-$0\secspt1130). The left panel of Figure \ref{fig_wcen_data} shows a smoothed X-ray three-color image of \wcen; the right panel displays a broad-band ($0.5-7.0$ keV) image of the inner 2\minspt7 $\times$ 2\minspt5 with proposed centers marked.

We applied CIAO's {\tt wavdetect} algorithm to events with energies in the range $0.5-7.0$ keV, and adopted a source significance threshold of 10$^{-6}$, which gives $\sim$1 false detection per 10$^6$ pixels \citep{Freeman02}, with wavelet (spatial) scales of 1--16 in intervals of 2. No X-ray point source is detected at or near any of the cluster's proposed centers; the closest X-ray detection lies $>$10\asec\ from the AvdM10 center (Fig. \ref{fig_wcen_data}).

In searching for an X-ray source associated with a possible IMBH, we allow that it might wander due to Brownian motion resulting from energy exchange with individual stars. The IMBH's ``wander radius'' is described by \citet{Chatterjee02} and scales as $<{\rm x}^2> = 2/9 ~{\rm r}_c^2 ~{\cal M}_{\star} / {\cal M}_{\rm BH}$, where ${\rm x}$ is the one-dimensional RMS offset from the cluster center,  \rc\ is the core radius, \mass$_{\star}$ is the average stellar mass, and \mass$_{\rm BH}$ is the mass of the IMBH. A less massive IMBH will experience larger perturbations so we conservatively adopt \mass$_{\rm BH} = 1000$ \msun\ (among the lowest IMBH mass limits for \wcen\footnote{This limit comes from the radio study of \citet{Lu11}, \emph{not} a dynamical study, since Brownian motion would wash out the peak in the velocity dispersion that provides dynamical evidence for an IMBH.}), and an average stellar mass of 0.8 \msun, which gives an RMS offset of $\sim 2$\asec. Hence, we search a circle with radius $\sim$ 6\asec\ (3$\sigma$) at each of the possible cluster centers (Fig. \ref{fig_wcen_data}).

We estimate the on-axis background in five large (45 pixel radius) source-free regions and find an average 0.5--7.0 keV background level of 0.0643 counts/pixel. In these high quality \Chandra\ data, on-axis sources are easily localized to within a 2-pixel radius ($\sim$ 1\asec), equivalent to an aperture area of 12.6 pix$^2$. With four possible cluster centers, each with a 6\asec\ search radius, we seek an IMBH in an area encompassing $\sim$1800 pixels. Given the very low background level, an on-axis source with just 6 counts in a 1\asec\ radius has a Poisson probability of $1.7\times10^{-4}$ and is likely to be real. Inspection of our search area confirms that none of the central regions contains a source with 6 or more counts.

\begin{deluxetable}{lll}
\tabletypesize{\scriptsize}
\tablewidth{0pt}
\tablecaption{\wcen\ Reference Parameters \& X-ray Detection Limits} 
\tablehead{\colhead{Description} & \colhead{Best Value} & \colhead{Ref.} }

\startdata
Mass                                 & $3\times10^6$ \msun               & 1   \\
Distance                             & 5.2 kpc                           & 2   \\
Half-mass radius                     & 4.2\amin $\sim$ 6.3~pc           & 2   \\
E(B-V)								 & 0.11                              & 3   \\
Galactic N$_H$                       & $1.2\times10^{21}$ cm$^{-2}$      & 4   \\
\\
Proposed Centers &                   \multicolumn{2}{l}{RA, Dec (J2000)}       \\
~~~AvdM10                            & 13:26:47.24,    $-$47:28:46.45    & 5   \\ 
~~~Noyola (2010)$^a$                 & 13:26:47.24,    $-$47:28:42.20    & 6   \\
~~~Noyola (2008)$^b$                 & 13:26:46.08,    $-$47:28:42.9     & 7   \\
~~~Harris (1996)                     & 13:26:45.9\phn, $-$47:28:36.9     & 2   \\
\\
PL photon index ($\Gamma$)           & 2.3 (M15, M32)                    & 8,9 \\                         
                                     & 2.7 (Sgr~A*)                      & 10   \\
\\
\multicolumn{3}{l}{X-ray detection limits (95\% confidence)$^c$} \\
~~~\fx($0.5-7.0$ keV)                & $\le 5.0 \times10^{-16}$ \ergscm2\ & 11   \\
~~~\lx($0.5-7.0$ keV)                & $\le 1.6 \times10^{30}$ \ergs\     & 11   \\
~~~\fx($0.5-2.5$ keV)                & $\le 2.8 \times10^{-16}$ \ergscm2\ & 11   \\
~~~\lx($0.5-2.5$ keV)                & $\le 9.0 \times10^{29}$ \ergs\     & 11   
\enddata
\tablecomments{$^a$The \citet{Noyola10} centroid has been tied to the master frame of AvdM10 at (x,y) = (6724,6895) and converted to RA, Dec using the 2MASS point-source catalog. $^b$The \citet{Noyola08} centroid has been corrected by AvdM10 for HST guidestar errors. $^c$The total combined \Chandra\ exposure time for these detection limits is 290.9 ks. The flux and luminosity limits are estimated for a PL with $\Gamma =  2.3$, and corrected for Galactic absorption (see \S\ref{obs} for details). Refs --- 1: \cite{Meylan02}; 2: \cite{Harris96}; 3: \cite{Lub02}; 4: \cite{Willingale13}; 5: AvdM10; 6: \cite{Noyola10}; 7: \cite{Noyola08}; 8: \cite{Ho03a}; 9: \cite{Ho03b}; 10: \cite{Baganoff03}; 11: This work.
}
\label{tab_wcen}
\end{deluxetable}

We use the {\tt aprates} tool with a 6\asec\ aperture at the AvdM10 cluster center to place an upper limit (95\% confidence) on the $0.5-7.0$ keV X-ray count rate of $4.43\times10^{-5}$ counts s$^{-1}$. An IMBH accreting at a low fraction of its Eddington luminosity should be well-described by an absorbed power-law (PL), the spectral model used in numerous IMBH studies \citep[e.g.,][]{Ho03a,Ho03b,MillerJones12}, and in modeling low-luminosity active galactic nuclei \citep[AGN, e.g.,][]{Dong12}. At a distance of 5.2 kpc and assuming a PL spectrum with $\Gamma = 2.3$ (Table \ref{tab_wcen}), this detection limit corresponds to an unabsorbed flux limit of \fx(0.5-7.0 keV) $\le 5.0 \times10^{-16}$ \ergscm2, or a luminosity limit of $\le 1.6 \times10^{30}$ \ergs\ (see Table \ref{tab_wcen} for $0.5-2.5$ keV limits). Adopting the best fit PL for Sgr~A* in quiescence \cite[$\Gamma = 2.7$,][]{Baganoff03} yields a nearly identical result. This is the lowest X-ray limit yet reported on an IMBH candidate in a globular cluster.

\section{Results and Discussion}
\label{results}

\begin{figure*}[th]
\begin{center}
\includegraphics[scale=0.55]{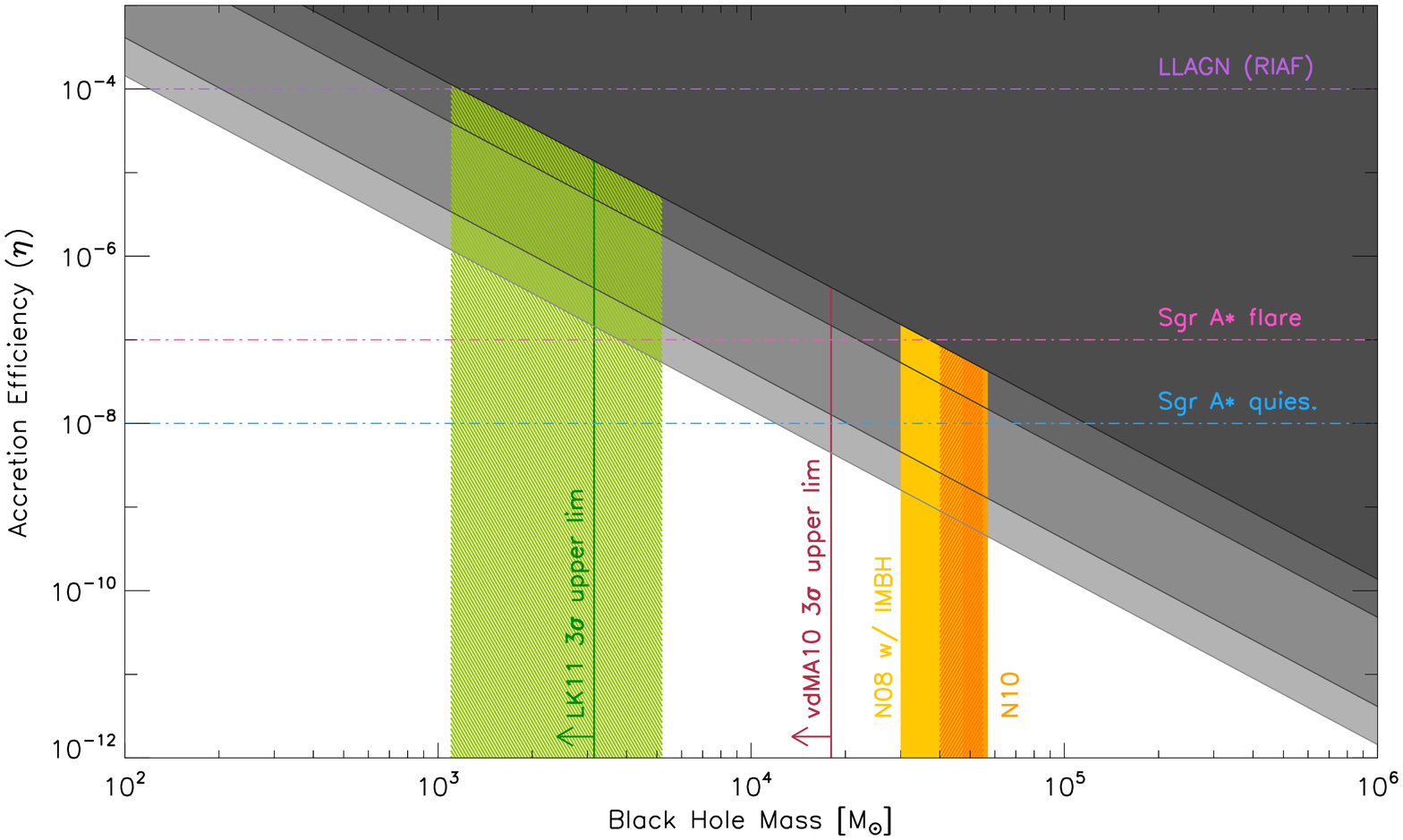}
\figcaption{Exclusion plot for an accreting IMBH in \wcen\ (mass of the black hole vs. accretion efficiency [$\eta$]) for our new X-ray upper limit, \lx($0.5-7.0$~keV) $\le 1.6\times10^{30}$ \ergs. The light (\lbol/\lx\ $= 7$) and medium (\lbol/\lx\ $= 20$) grey regions are excluded for a simple Bondi accretion scenario (Eqns. 2 \& 3). The medium-dark and dark grey regions are excluded for the modified accretion description from \citet{Maccarone08}, discussed in \S \ref{accretion_lims} (Eqn. 4). The gas density and temperature are taken to be $n \sim 0.038$ cm$^{-3}$ and $T=10^4$ K, respectively (\S\ref{accretion_lims}). The colored vertical lines/polygons correspond to IMBH mass estimates or upper limits (with errors) from radio and dynamical studies: 3$\sigma$ radio upper limit from ATCA \citep[best fit: dark green line, error interval: light green polygon,][]{Lu11}; 3$\sigma$ upper limit from vdMA10, including both core and cusp dynamical models (dark red line; see \S\ref{mass_lims}); mass range from a combined \HST\ and IFU dynamical study \citep[yellow polygon,][]{Noyola08}; revised \HST\ plus IFU study for new (light orange polygon) and old (hashed dark orange polygon) dynamical centers \citep{Noyola10}. Horizontal dot-dashed lines mark efficiencies for various accreting black holes (see also Table 2): low-luminosity AGN (LLAGN with $\eta \sim 10^{-4}$; purple); Sgr~A* (\mass$_{\rm BH} = 4.1\times 10^6$\msun), during a flare \citep[magenta,][]{Nowak12,Neilsen13} and in quiescence \citep[blue,][]{Baganoff03}. An AGN (\mass$_{\rm BH} > 10^6$\msun) with a ``canonical'' thin-disk efficiency ($\eta = 0.1$) would lie two decades above the top of this plot and is excluded for the entire IMBH mass range. \\
(A color version of this figure is available in the online journal.)}
\end{center}
\label{fig_wcen_exclusion}
\end{figure*}

\subsection{Eddington Ratio and Accretion Luminosity}
\label{accretion_lims}

We convert this \lx\ limit to a bolometric luminosity limit via the standard correction \citep[$\sim7-20$,][]{Elvis94,Ho03a}, \lbol\ $< (1.1-3.2)\times10^{31}$ \ergs. The Eddington luminosity,  
\begin{equation}
  {L}_{\rm Edd} = 1.26\times10^{38} ~({\cal M}_{\rm BH}/{\cal M}_{\odot}) ~{\rm erg~s^{-1}} \\
\end{equation}
for the 3$\sigma$ upper limit on the IMBH mass from vdMA10 ($\le 1.8 \times 10^4$ \msun) is ${L}_{\rm Edd} = 2.3\times10^{42} ~{\rm erg~s^{-1}}$. This implies a very low Eddington accretion ratio of \lx/\ledd\ $< 7.0\times10^{-13}$ or \lbol/\ledd\ $< (5-14)\times10^{-12}$.

For the simplest case of spherical accretion \citep{Bondi52}, the accretion rate and luminosity (scaled to quantities appropriate for \wcen) can be estimated from an optically thick, geometrically thin disk \citep{Shakura73},
\begin{equation}
  \begin{aligned}
\dot{\cal M}_{\rm Bondi} & = 1.37 \times 10^{-8}\left({{{\cal M}_{\rm BH}} \over {10^4 ~{\cal M}_{\odot}}}\right)^2 \\ 
      & \times \left({{n} \over {0.038 ~{\rm cm^{-3}}}}\right) \left({T} \over {10^4 ~\rm{K}}\right)^{-3/2} {\cal M}_{\odot}~{\rm yr^{-1}}, ~{\rm and}
  \end{aligned}
\end{equation}
\begin{equation}
L_{\rm acc,1} = c^2~\eta~\dot{\cal M}_{\rm Bondi}, 
\end{equation}
where $\eta$ is the accretion efficiency \citep{Ho03a}. We estimate the core gas density following the formalism in \cite{Pfahl01}, based on free expansion of mass lost from stars, $n = 1\left({\cal M}_{\rm{c}} \over 10^5 {\cal M}_{\odot}\right)\left(v_w \over 20~{\rm km~s^{-1}}\right)^{-1} \left(r_{\rm{\star}} \over 0.5~{\rm pc}\right)^{-2}$ cm$^{-3}$. We take the enclosed mass ${\cal M}_{\rm{c}} = {\cal M}_{\rm{h}}$ (i.e., half the cluster's mass), $r_{\rm{\star}}$ = $r_{\rm{h}}$ (the half-mass radius, $\sim 1.6~r_{\rm{c}}$), and a characteristic wind speed $v_w =$ 50 km s$^{-1}$, and find $n \sim 0.038$ cm$^{-3}$, not unlike what \citet{Freire01a} measure for 47 Tuc. The gas temperature is assumed to be $T= 10^4$ K. For ${\cal M}_{\rm BH} = 1.8 \times 10^4$ \msun, $L_{\rm acc,1} = \eta ~2.5\times10^{39}$ \ergs. Comparing this accretion luminosity to the bolometric luminosity, we find $\eta \sim (4-13) \times 10^{-9}$. In this scenario, if \wcen\ hosts such a massive IMBH, it must be a \emph{very} inefficient accretor.

\citet{Maccarone08} have developed a more conservative approach for estimating IMBH accretion efficiencies and luminosities. They assume that (1) the accretion rate (\mdot) is $\sim$3\% of the Bondi rate for gas at $T = 10^4$ K \citep{Pellegrini05}; (2) the system is below the ``low-hard state'' transition \citep[e.g.,][]{Maccarone03}, wherein \mdot/\mdot$_{\rm{Edd}} < 0.02$ for \mdot$_{\rm{Edd}} = $ \ledd/(0.1$~c^2$); and (3) $\eta$ scales linearly with accretion rate --- to insure continuity at the transition ($\sim0.1$), $\eta = 0.1 \left((\dot{\cal M}/\dot{\cal M}_{Edd}) / 0.02 \right)$. The accretion luminosity is then
\begin{equation}
L_{\rm acc,2} = c^2~4.5\times10^{-3}\left(\dot{\cal M}_{\rm{Bondi}}^2/\dot{\cal M}_{\rm{Edd}}\right), 
\end{equation}
or $L_{\rm acc,2} \sim 1.3 \times 10^{33} ~{\rm erg~s^{-1}}$, where ${\cal M}_{\rm BH} = 1.8 \times 10^4$ \msun, $\dot{\cal M}_{\rm{Edd}} = 4.0\times10^{-4}$ \msun\ yr$^{-1}$, and $\eta = 1.7\times10^{-5}$.

\subsection{Constraints on an IMBH}
\label{mass_lims}

A simple test for the existence of a $\sim1.8\times10^4$ \msun\ IMBH in \wcen\ assumes a standard thin-disk accretion efficiency $\eta \sim 0.1$; the resulting accretion luminosity is $\sim2.5\times10^{38}$ \ergs. No source this bright is present in \wcen. The brightest X-ray sources are two known CVs \citep[\lx\ $= 1.4-1.5 \times10^{32}$ \ergs,][]{Haggard09,Cool13}, both of which have optical counterparts and \lx\ several orders of magnitude below this hypothetical accreting IMBH. 

We can convert the \lx\ (or \lbol) upper limit to an IMBH mass upper limit by assuming a model-dependent accretion efficiency for a radiatively inefficient accretion flow \citep[RIAF, e.g.,][and references therein]{Yuan03,Narayan08} with a low accretion efficiency ($\eta \sim10^{-4}$). For $L_{\rm acc,1} =$ \lbol, Eqns. 2 and 3 then imply \mass$_{BH}$ \simlt\ $(120-200)$ \msun. The more conservative accretion scenario that informs Eqn. 4 supports a similarly low accretion efficiency, and a comparison between $L_{\rm acc,2}$ and \lbol\ implies \mass$_{BH}$ \simlt\ $(3.7-5.3)\times10^{3}$ \msun. Even in the conservative case, this is an order of magnitude below  dynamical estimates.

Figure 2 summarizes these findings as an exclusion plot based on the black hole mass and the accretion efficiency (Eqns. 1--4). Our 0.5--7.0 keV X-ray luminosity limit rules out the grey regions, which represent the scenarios described in \S3.1 --- those combinations of \mass$_{\rm BH}$ and $\eta$ would result in an observable X-ray flux. IMBH mass limits (or expectations) from radio and dynamical studies are shown as colored polygons; efficiencies for several known massive black hole systems are indicated for reference. 
An IMBH in the expected range for \wcen\ ($3\times10^{3}-3\times10^{4}$ \msun) must be accreting with very low efficiency ($\eta < 10^{-5}-10^{-9}$) to fall below this low X-ray limit. A scenario in which \wcen\ does not host an IMBH cannot be ruled out and, under most thin-disk and RIAF models, this scenario is preferred. The gas density and temperature of the accretion flow are based on theoretical predictions (\S3.1) --- a measure of the conditions in \wcen's ambient medium would help to tighten these constraints.

\subsection{The Fundamental Plane}
\label{FP}

The Fundamental Plane (FP) for black hole activity is an observed correlation between radio luminosity ($L_{\rm R}$), X-ray luminosity, and black hole mass \citep{Merloni03,Falcke04}, designed for estimating $M_{BH}$ for given $L_{\rm R}$ and \lx. \citet{MillerJones12} give a recent formulation of the FP in a study of the cluster G1 in M31,
\begin{equation}
  \begin{aligned}
{\rm log}~&{\cal M}_{\rm BH} = (1.638 \pm 0.070) ~{\rm log}~L_{\rm R} \\
                                      & - (1.136 \pm 0.077) ~{\rm log}~L_{\rm X} - (6.863 \pm 0.790).
  \end{aligned}
\end{equation}

With only upper limits on \lx\ and $L_{\rm R}$ in \wcen, application of the FP is not likely to be valid. If we conclude that there is no true radio detection, the FP provides no information. If instead we assume that the 2.5$\sigma$ radio detection from \citet[][$f_{\rm R} = 17.5 \mu{\rm Jy}$ at 5.5 GHz]{Lu11} is real, we find $L_{\rm R} \sim 1.1\times10^{27}$ \ergs. Combining this with our X-ray limit\footnote{Eqn. 5 requires \lx\ in the $0.5-10$ keV band; we convert our $0.5-7$ keV limit using PIMMS and the parameters in Tab. 1 to find \lx$(0.5-10~{\rm keV}) < 1.8\times10^{30}$ \ergs.} would produce a \emph{lower} limit on the IMBH mass of \mass$_{BH}$ \simgt\ $1.2\times10^3$ \msun. This is in contrast to the IMBH mass upper limit derived for simple Bondi accretion (Eqns. 2 and 3), but not inconsistent with the mass upper limit estimated for the modified Bondi scenario (Eqn. 4). Hence, if the radio detection is real, the accretion source may fall on the FP. However, we conclude that the FP does not actually help constrain the mass of any IMBH in \wcen, given the current observations.

\subsection{\wcen\ in Context}

\begin{deluxetable}{lcccc}
\tabletypesize{\scriptsize}
\tablewidth{0pt}
\tablecaption{Black Holes Exhibiting Low-Efficiency Accretion} 
\tablehead{\colhead{} & \colhead{${\cal M}_{\rm BH}$}  & \colhead{\lx} & \colhead{} & \colhead{}  \\
           \colhead{Target} & \colhead{(\msun)} & \colhead{(erg s$^{-1}$)}  & \colhead{\lx/\ledd} & \colhead{Ref.} }
\startdata
\multicolumn{5}{l}{{\bf Proposed GC IMBHs}}\\
\wcen     & $< 1.8$\x$10^4$  & $<1.6$\x$10^{30}$  & $<7.0$\x$10^{-13}$    & 1,2\\ 
M54       & $\sim9.4$\x$10^3$& $<1.5$\x$10^{32}$  & $<1.4$\x$10^{-10}$    & 3,4\\ 
M15       & $< 2000$	     & $<5.6$\x$10^{32}$  & $<2.2$\x$10^{-9}$     & 5,6\\ 
\cline{1-5}
\multicolumn{5}{l}{{\bf Known Accreting Black Holes}}\\
Sgr~A*(q) & $4.1$\x$10^6$  & $\sim2.4$\x$10^{33}$ & $5$\x$10^{-12}$       & 7 \\ 
Sgr~A*(f) & $4.1$\x$10^6$  & $(1-20)$\x$10^{34}$  & $(2-40)$\x$10^{-11}$  & 7,8,9\\ 
M32       & $2.5$\x$10^6$  & 9.4\x$10^{35}$       & $3$\x$10^{-9}$        & 10
\enddata
\tablecomments{For Sgr~A* we include both (q: quiescent) and (f: flare) X-ray properties. Refs --- 1: This work; 2: vdMA10; 3: \citet{Ibata09}; 4: \citet{Wrobel11}; 5: \citet{Ho03a}; 6: \citet{Strader12}; 7: \citet{Baganoff03}; 8: \citet{Nowak12}; 9: \citet{Neilsen13}; 10: \citet{Ho03b}.
}
\label{tab_eta_compare}
\end{deluxetable}

A similar dearth of IMBH accretion signatures has been noted in several other GCs, including M15, M19, and M22 \citep{Strader12}, and the above-mentioned case of G1 \citep{MillerJones12}. The dynamical measures that have led to claims for IMBHs \citep{Ulvestad07,Noyola08,Noyola10}, as well as in NGC 6388 \citep{Lutzgendorf11}, and M54 \citep{Ibata09}, remain controversial and may be in error \citep[radio non-detections have also been reported for NGC 6388 and M54,][]{Cseh10,Wrobel11}, or simply suffer from large uncertainties. If the dynamical measures are not in error, we may be probing an unfamiliar accretion regime, in which the ambient gas is extremely sparse, or the accretion is very inefficient.

A handful of black holes are known to accrete with very low X-ray efficiencies; we include several examples in Table \ref{tab_eta_compare}. For a massive IMBH in \wcen\ ($1.8 \times 10^4$\msun), \lx/\ledd\ would be the lowest even in this poorly understood regime. Of course, the IMBH in \wcen\ may be less massive or may not be present, leaving an alternate question: Why hasn't an IMBH formed at the dynamical center of this dense stellar system?

\section{Summary}
\label{summary}
 
We find no evidence for an X-ray source associated with \wcen's cluster center in a deep ($\sim291$ ks) \Chandra\ co-add. We report an X-ray upper limit for the (unabsorbed) flux and luminosity of \fx$(0.5-7.0~{\rm keV}) < 5.0\times10^{-16}$ \ergscm2, and \lx$(0.5-7.0~{\rm keV}) < 1.6\times10^{30}$ \ergs, respectively. If an IMBH with \mass$_{\rm BH} \sim 1.8 \times 10^4$\msun\ resides in \wcen, it has the lowest Eddington ratio of any known massive BH, even lower than Sgr~A* in quiescence. 

\acknowledgments

We appreciate discussions with Nicolas Cowan and Fred Rasio and thank the referee for helpful comments that improved the manuscript. COH is supported by an NSERC Discovery Grant and an Ingenuity New Faculty Award. This work is supported by \Chandra\ Award Numbers GO2-13057A and GO2-13057B issued by the CXO, 
which is operated by the Smithsonian Astrophysical Observatory for and on behalf of NASA 
under contract NAS8-03060. This research made use of data obtained from the \Chandra\ Data Archive and software provided by the \Chandra\ X-ray Center (CXC) in the application packages CIAO, ChIPS, and Sherpa.

{\it Facility}: \facility{CXO (ACIS)}


\end{document}